\documentclass[11pt]{article}
\usepackage[utf8]{inputenc}
\usepackage{graphicx}
\usepackage{verbatim}
\usepackage{amsmath}
\usepackage{fancyvrb}
\usepackage{xcolor} 
\usepackage{overpic}
\usepackage{subcaption}
\usepackage{placeins}
\usepackage{authblk}
\usepackage{soul}

\usepackage{fullpage}

\usepackage{ulem}
\usepackage{enumitem}

\usepackage{tikz}
\usetikzlibrary{positioning}
\usetikzlibrary{arrows.meta}
\usetikzlibrary{shapes.geometric}

\title{Hypergraph Topological Features for  Autoencoder-Based Intrusion Detection for Cybersecurity Data}

\author[1]{Bill Kay}
\author[1]{Sinan G. Aksoy}
\author[1]{Molly Baird}
\author[1]{Daniel M. Best}
\author[1]{Helen Jenne}
\author[1]{Cliff Joslyn}
\author[1,2]{Cristopher Potvin}
\author[1]{Gregory Henselman-Petrusek}
\author[1]{Garret Seppala}
\author[1]{Stephen J. Young}
\author[1]{Emilie Purvine}
\affil[1]{Pacific Northwest National Laboratory, Richland, Washington, USA}
\affil[2]{Department of Mathematics, Michigan State University, East Lansing, Michigan, USA}

\begin{document}








\maketitle

\begin{abstract}
  In this position paper, we argue that when hypergraphs are used to capture multi-way local relations of data, their resulting  topological features describe global behaviour. Consequently, these features capture complex correlations that can then serve as high fidelity inputs to autoencoder-driven anomaly detection pipelines. We propose two such potential pipelines for cybersecurity data, one that uses an autoencoder directly to determine network intrusions, and one that de-noises input data for a persistent homology  system, PHANTOM. We provide heuristic justification for the use of the methods described therein for an intrusion detection pipeline for cyber data. We conclude by showing a small example over synthetic cyber attack data. 

\end{abstract}

  \twocolumn   
\section{Introduction}
\label{sec:cyber}

Cybersecurity is the protection and recovery of computing devices, networks, and data from adverse effects~\cite{NISTFramework}. This can  distilled into three components: confidentiality, integrity, and availability (CIA)~\cite{AndersonJ1972}. Confidentiality states that only authorized users have access to information and systems, integrity states that data has not been tampered with, and availability means users have timely, reliable access to information~\cite{NIST800-12}.  Cybersecurity experts must protect  the CIA triad from  malicious and inadvertent actors. 

While the overall concept of cybersecurity is understood, effectively facilitating cybersecurity is complicated given the massive amounts of data being captured, the complex nature of systems and applications, the continual advancement of adversary techniques, and the general lack of security understanding by most computer users.  Cybersecurity defenders are tasked with protecting networks and systems. As defenses grow in complexity, so too do adversarial tactics. This escalation results in the collection of increasing amounts of log data from network appliances and computer protection applications to try and stay ahead of vulnerabilities and block or mitigate attacks.  It is not feasible for cybersecurity defenders to review and search all this data. Instead  they are dependent on detection alerts, indicators, and blocking to provide pointers for  investigation and to stop attacks before execution.

Early cybersecurity detection centered around the development of signatures for attacks, scripts, and code.  Known malicious code was analyzed to develop signatures, with selected portions of the known malicious binary used to check for matches with a suspected binary. The selected portions could include the order of operating system calls, unique strings, or other static indicators.  The issue with signature-based detection is that it is easily circumvented through means like  binary encoding, renaming variables, and encryption.  New techniques such as behavioral algorithms, machine learning, and artificial intelligence provide detections that are robust to binary alteration techniques.

To support anomaly detection in cybersecurity data, and with the recognition that attacker behavior has become more complex and subtle, we propose a machine learning (ML) workflow based on topological measurements of hypergraph-structured cyber data. In Section~\ref{sec:hypergraph}, we discuss the utility of hypergraph data representations. In Section~\ref{sec:topology} we detail how to derive powerful topological network descriptors from such a hypergraph data representation. In Section~\ref{sec:autoencoder} we introduce an autoencoder, a particular type of neural network suitable for anomaly detection and denoising of data. We describe how autoencoders can be applied directly to topological data from time series of hypergraphs to detect network intrusion in Subsection~\ref{sec:aedirect}. We introduce the PHANTOM framework for topological anomaly detection in Subsection~\ref{sec:phantom} and discuss how ML  can be used as a pre-processing step to provide robustness to noisy input data. 
Finally, in Section~\ref{sec:ARC} we work though a small topology-based hypergraph example on real world data.

We posit that (1). Complex interrelations in cyber data are well-modeled by a hypergraph framework; (2). Topological measures of hypergraphs capture both global and local structural properties; and (3). These signatures can be used in an ML framework, such as an autoencoder, to aid anomaly detection efforts in cybersecurity.

\section{Hypergraphs}
\label{sec:hypergraph}
\begin{figure}[h]
     \centering
     \includegraphics[width=1\linewidth]{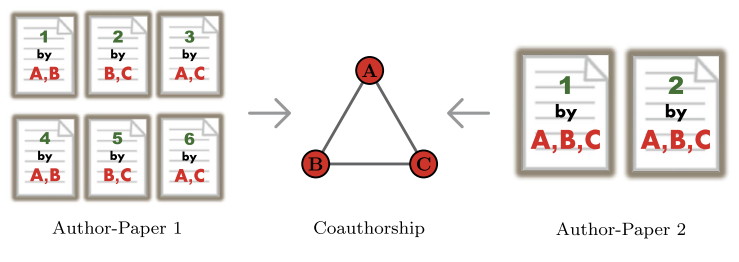}
     \caption{Two sets of papers that yield the same coauthorship graph but different coauthorship hypergraphs~\cite{aksoy2020hypernetwork}.}
     \label{Fig:Loss}
\end{figure}

A \ul{hypergraph} $H= (V, E)$ is an ordered pair, where $V$ is a vertex set and 
set $E$ such that every edge $e \in E$ is a subset of $V$ (not necessarily all the same size)
In the case that all edges are of size  $2$, the hypergraph is just a graph. 
While graphs represent pairwise relationships, hypergraphs model  relationships of any number of elements, including  more than two, or even singletons.
Graphs are a natural model of many systems (e.g., a social network where friendship relationships are symmetric), but often times multiway relational data cannot be accurately modeled by a graph without losing information, as illustrated in Figure~\ref{Fig:Loss}.

Both graphs and hypergraphs are  defined by local relationships that give rise to global structure. While graphs provide a compressed (albeit lossy \cite{kirkland2018two}) picture of data that contains multi-way relations, hypergraphs provide a more accurate picture at the cost of additional (sometimes intractable) computation. 
The additional information captured by a hypergraph is reflected in its global properties, and these properties reflect meaningful features of the underlying data. For example, given a time series of hypergraphs constructed from network data, drastic changes in  hypergraph properties  can correlate with big shifts in network activity, perhaps due to anomalous or intrusive behaviour. Hypergraph structures used to model cyber data vary between use cases, the nature of the data, and the problem at hand. We will not describe all possible hypergraph constructions from cyber data here, but rather assume that  a hypergraph representation of our data has been provided. For one illustrative example of how to meaningfully construct a hypergraph  explicitly from real life cyber data, see Section~\ref{sec:ARC}. While this data arose from red team activity in a small test system, the anomaly detector that we posit does require access to both clean and dirty data. We do not posit a specific solution to this problem, but acknowledge that this is a challenge facing machine learning based anomaly detection.
Rather than concentrating on traditional hypergraph measures that could be used for anomaly detection (degree distribution, connectivity, etc., for a full survey, see~\cite{aksoy2020hypernetwork,berge1984hypergraphs}), we will focus on those features derived from combinatorial topology which we overview in Section~\ref{sec:topology}.

\section{Topology}
\label{sec:topology}

{\em Algebraic topology}
leverages algebra and combinatorics to study shapes that are otherwise hard to understand, e.g.\ because they are high dimensional, lack a closed-form description, or lack a well-formed distance metric. 
The primary objects of interest are topological spaces and continuous maps between them.  


To capture information about a hypergraph topologically, we first transform it to a topological space called a \emph{simplicial complex}.  A simplicial complex consists of a collection of vertices (0-simplices), edges (1-simplices), triangles (2-simplices), tetrahedra, etc.  Figure \ref{Fig:SP}(c), (d) shows two examples.  There are several ways to construct simplicial complexes from a  hypergraph~\cite{hom_gr_hyp,horak2013spectra}, and
we describe the {\em restricted barycentric subdivision} in Section~\ref{sec:ARC}. See~\cite{bressan2019embedded,grigoryan_homology_2019,muranov2021path} for further topological techniques.

The shape of simplicial complexes can be analyzed via {\em homology}. Homology reveals spatial properties informally known as ``holes'', which are interpreted as connected components (dim-0 holes), loops (dim-1 holes), and  voids (dim-$k$ holes). Formally, the homology of a topological space is a sequence of vector spaces.  The dimension of the $k^{th}$ space in the sequence is called the $k^{th}$ {\em Betti number}, and represents the number of dim-$k$ holes. 

Betti numbers can be used to identify structural changes in time-varying data: if we build a hypergraph to represent the network flow data, then the associated Betti numbers represent the numberof  voids in data.  Large changes in these numbers over time  may indicate that network activity has changed qualitatively -- either by ``filling in'' voids -- possibly  due to a breach in network security resulting in data flow between nodes that typically do not communicate, or nodes communicating more often than  usual -- or by opening new voids. 
Thus, deviation from baseline Betti numbers could be used to signal anomalous behavior.

Closely related to homology is the \ul{$k$-Hodge Laplacian}, a matrix whose eigenvalues encode random walk information about a simplicial complex. For example, the $0$-Hodge Laplacian is the celebrated {\em combinatorial Laplacian}~\cite{chung1997spectral} whose spectrum is tied to  random walks; it provides a powerful technique for sampling  network activity.   

Another important topological measurement of data is \ul{persistent homology} (PH) \cite{edelsbrunner2000topological,zomorodian2005computing}, which measures the shape of a point cloud (i.e., a collection of high dimensional numerical vectors) at multiple ``scales,'' and relates these measurements together.  
From point cloud data, PH builds an increasing sequence of simplicial complexes by introducing a distance parameter, $\epsilon$. 
For each $\epsilon$ a simplicial complex is constructed by connecting all pairs of points that are within distance $\epsilon$ of each other, and all $k$-cliques of the resulting graph structure are filled in with $k$-simplices (this is known as the Vietoris-Rips complex).
Starting with the point cloud itself at $\epsilon=0$, as $\epsilon$ increases the simplicial complex takes shape; higher dimensional voids emerge (are ``born''), and later are filled in (``die'').
The birth and death distance thresholds for each of the topological features determine intervals of $\epsilon$ which form a fingerprint called a \emph{barcode} or \emph{persistence diagram} that encodes information about the overall shape of a manifold from which the point cloud could have been sampled from.

These topological measures provide a suite of computational tools to recover global high dimensional analogues of familiar combinatorial properties from hypergraphs and point cloud data. Further, these properties will often have hidden correlations that are hard to detect by inspection. Machine Learning (ML) provides a general framework for detecting complex correlations in data. 

\section{Autoencoders as Anomaly Detectors}
\label{sec:autoencoder}

An \ul{autoencoder} is a particular neural network architecture that, loosely speaking, learns efficient encodings of (often high dimensional) data~\cite{kramer1991nonlinear}. The two primary components are an encoder which learns a low dimensional encoding of the input, and a decoder which learns an approximation of the original input from the low dimensional embedding. Importantly, one internal layer should be strictly smaller than the input or output, serving as a bottleneck.
This smaller internal layer can be used as a latent, or low-dimensional, representation of the input data.

\subsection{Direct method of anomaly detection}
\label{sec:aedirect}
Autoencoders can be used to directly build anomaly detectors by training on data known to be normal with a prescribed loss function between the input and the output. 
Training stops when outputs are consistently close to the input (according to a prescribed threshold). The autoencoder is then deployed on unlabeled data, and if outputs then start to vary significantly from inputs, that provides evidence that an anomaly has occurred (that is, the data are behaving differently from the normal training data). Often times, the input is a sequence of vectorized values (e.g., bag of words, tf-idf, or perhaps Betti numbers from hypergraphs) from which the autoencoder can learn complex correlations that are difficult to identify by inspection~\cite{golczynski2021end,hanselmann2020canet}. 

\subsection{Autoencoders in PH-based Anomaly Detection}
\label{sec:phantom}

As described in Subsection~\ref{sec:aedirect}, autoencoders trained on vectorized representations such as topological features of hypergraphs derived from cyber data can provide a direct solution to the problem of anomaly detection for data such as network flow. Such an approach can fail when normal data have high variability (e.g., large fluctuations due to noise) making it  be difficult to distinguish said noise from anomalies. 
Autoencoders are adept at denoising data~\cite{zurich2018deep}, and so as a second approach we propose an autoencoder as an ML processing step for our existing persistent homology based anomaly detector, PHANTOM (Persistent Homology ANomalous Tracking Observation Monitor)\footnote{When released: https://github.com/pnnl/PHANTOM} \cite{bruillard_anomaly_2016}. 

PHANTOM reads log files in which each log has a timestamp and a set of fields (e.g., source IP, user, hostname, number of bytes) and partitions data into time windows of uniform length. A collection of user-defined statistics (e.g., the average of the ``bytes'' field, or the number of unique IPs seen) is computed for each window to create a high dimensional vector summary of the data in the time window. These vectors are computed for a set of data known to be anomaly free to establish a baseline. We compute the PH barcode for the point cloud of vectors comprising the baseline to establish an initial topological signature of ``normal.'' As more data are collected, a summary vector is computed for each new time window, which is then added to the baseline for which the PH barcode is recomputed. We compare the barcode of the baseline plus the current observation to the barcode of just the baseline using the Wasserstein distance~\cite{edelsbrunner2022computational} (as implemented in the persim Python package) to establish an anomaly score for the new time window. If this Wasserstein anomaly score is small the new data point is determined to be non-anomalous and the baseline is updated with this new point replacing the oldest point (and the barcode is recomputed). However, if it is ``Wasserstein-far'' from the baseline according to a prescribed threshold the data point is flagged as anomalous and not added to the baseline. 
Not only has PHANTOM demonstrated efficacy in detecting the presence of anomalies, but the vector entry that causes the new time window to be far from the baseline can give insight into \textit{where} the anomaly is occurring. 

Presently, PHANTOM's vectorized inputs are user-defined for each use case. We are working to  incorporate summary properties of hypergraphs, including both basic statistics and the topological features described in Section~\ref{sec:topology}, into PHANTOM. We additionally expect that given the noted successes of autoencoder methods to summarize data into a low dimensional representation, an antoencoder trained on either the raw log data within a time window or a topological feature vector derived from a time window, will uncover previously unseen types of anomalies.
Hence, employing an autoencoder in the pipeline in this way could make PHANTOM robust to noisy data.

\section{Use Case: Evolving Hypergraph Topology}
\label{sec:ARC}

In this section, we will build a hypergraph from realistic cyber data generated on a test range, compute a topological property of this hypergraph (maximum degree) and argue heuristically that this property is a good indicator of a particular kind of anomaly (a port scan). While this is not a sophisticated use-case, it suffices for our position paper to suggest that more advanced hypergraph analysis could detect more sophisticated anomalies.


Our deception data set was created by the Pacific Northwest National Laboratory (PNNL) Asymmetric Resilient Cybersecurity (ARC) Initiative. 
Its network flow data set includes both normal activity and red team adversarial activities. 
Each record includes its start time (\verb+sTime+), end time (\verb+eTime+), source IP address (\verb+sIP+), destination IP address (\verb+dIP+), source port (\verb+sPort+), destination port (\verb+dPort+), and packet header (\verb+packet+) flags \cite{silk2022}.

These data are used to construct a time series of hypergraphs, indexed by time ranges. For each fixed time window, we take each flow whose \verb+sTime+ falls in the window. While SiLK flows are typically unidirectional, we consider the bidirectional pairing of two (or more) records to make up a single record for analysis. This creates a record where the source is the origin (client) and the destination is recipient (server) of the full network communication. 

\begin{figure}[t]
    \centering
    \includegraphics[width=1\linewidth]{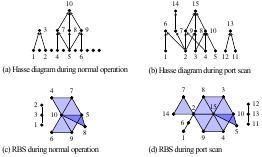}
    \caption{Normal vs. anomalous activity }
        \vspace{-.5cm}
    \label{Fig:SP}
\end{figure}

The vertex set of our hypergraph is the set of all source IPs and the edges  are labeled by destination ports. 
A source IP vertex is included in a destination port edge if and only if that IP accessed that port during the given time window.
Note that the IP-port relationship is multi-way, in that an IP can access many ports during a period of time, and thus naturally determines a hypergraph.

Our hypergraph contains all the communication information, and is likely to be sensitive to anomalous behaviour. One  known attack phenomenon in the ARC network flow data is a scan across the network for a large range of ports. Any IP performing such a scan will therefore be contained in many hyperedges (one for each port scanned), and by the nature of the event most of those ports will only be used by the scanner while some more common ports will be used by other IPs. We expect that there will be an increased level of hyperedge containment during a port scan. 

In order to measure this we construct the hypergraph's edge containment partial order (ECP), the first step towards its {\em restricted barycentric subdivision} (RBS) \cite{hom_gr_hyp}.
The ECP is a directed graph in which the vertices are the hyperedges of the hypergraph, and there is an edge from hyperedge $e$ to hyperedge $f$ if $e\subseteq f$. Thus the in-degree (resp. out-degree) of a vertex in the ECP gives the number of edges it contains (resp. is contained in) in the hypergraph. 
The full RBS is the \emph{order complex} of the ECP which is a simplicial complex formed by replacing each chain $e_{i_0}\rightarrow e_{i_1} \rightarrow \cdots e_{i_k}$ in the ECP with a $k$-simplex.
Figure ~\ref{Fig:SP} shows both
the Hasse diagram (removing transitive links) of the ECP and the RBS of hypergraphs during a period of normal activity and during a scanning attack.
The maximum in-degree and out-degree of the ECP during the period of normal activity are 6 and 4, respectively, while the maximum in-degree and out-degree of the ECP during the scan are 7 and 6, respectively. 
Empirically, this phenomenon seems common, and so we will choose maximum degree of the ECP of a network hypergraph as one of the entries of a vectorized ML model described in Subsections~\ref{sec:aedirect} and~\ref{sec:phantom} to be sensitive to scan attacks. 


Our current work is focused on understanding the hypergraph conditions that give rise to, and destroy,  topological features in the RBS. As we grow this understanding we will identify more complex adversary techniques related to these topological phenomena. Namely, for future operations we hope to demonstrate the detection of novel adversarial attacks which have yet to be characterized. Topological features such as higher order homology and hypergraph spectra can change drastically when hypergraph structure is modified in ways that are difficult to quantify, and are thus prime candidates for anomaly detection measures when the nature of an attack is not known \textit{a priori.} We also note that while finding the maximum degree in a hypergraph is not much more difficult than finding the maximum degree in a graph, some more advanced topological statistics have significant computational overhead. Our future work will include an effort to reduce computational costs on hypergraph data known to arise from certain real-world sources based on known features (edge containment, sparsity, etc.).

\bibliographystyle{plain}
\bibliography{bib}


\end{document}